\documentclass[twocolumn,prb,aps,superscriptaddress,showpacs]{revtex4-1}

\usepackage{amsmath,amssymb}
\usepackage{graphicx}
\usepackage{xcolor}
\usepackage{ulem}

\begin{document}

\title{Fermi surface instabilities in CeRh$_{2}$Si$_{2}$ at high magnetic field and pressure}

\author{A. Palacio Morales}
\affiliation{Univ. Grenoble Alpes, INAC-SPSMS, F-38000 Grenoble, France}
\affiliation{CEA, INAC-SPSMS, F-38000 Grenoble, France}
\author{A. Pourret}
\email[E-mail me at: ]{alexandre.pourret@cea.fr}
\affiliation{Univ. Grenoble Alpes, INAC-SPSMS, F-38000 Grenoble, France}
\affiliation{CEA, INAC-SPSMS, F-38000 Grenoble, France}
\author{G. Seyfarth}
\affiliation{Univ. Grenoble Alpes, LNCMI, F-38042 Grenoble Cedex 9, France}
\affiliation{CNRS, Laboratoire National des Champs Magn\'etiques Intenses LNCMI (UJF, UPS, INSA), UPR 3228, F-38042 Grenoble Cedex 9, France}
\author{M.-T. Suzuki} 
\affiliation{RIKEN Center for Emergent Matter Science, Hirosawa 2-1, Wako, Saitama 351-0198, Japan}
\author{D. Braithwaite}
\affiliation{Univ. Grenoble Alpes, INAC-SPSMS, F-38000 Grenoble, France}
\affiliation{CEA, INAC-SPSMS, F-38000 Grenoble, France}
\author{G. Knebel}
\affiliation{Univ. Grenoble Alpes, INAC-SPSMS, F-38000 Grenoble, France}
\affiliation{CEA, INAC-SPSMS, F-38000 Grenoble, France}
\author{D. Aoki }
\affiliation{Univ. Grenoble Alpes, INAC-SPSMS, F-38000 Grenoble, France}
\affiliation{CEA, INAC-SPSMS, F-38000 Grenoble, France}
\affiliation{Institute for Materials Research, Tohoku University, Oarai, Ibaraki, 311-1313, Japan}
\author{J. Flouquet}
\affiliation{Univ. Grenoble Alpes, INAC-SPSMS, F-38000 Grenoble, France}
\affiliation{CEA, INAC-SPSMS, F-38000 Grenoble, France}
\date{\today } 

\begin{abstract}
We present thermoelectric power (TEP) studies under pressure and high magnetic field in the antiferromagnet CeRh$_{2}$Si$_{2}$ at low temperature. Under magnetic field, large quantum oscillations are observed in the TEP, $S(H)$, in the antiferromagnetic phase. They suddenly disappear when entering in the polarized paramagnetic (PPM) state at $H_{c}$ pointing out an important reconstruction of the Fermi surface (FS). Under pressure, $S/T$ increases strongly of at low temperature near the critical pressure $P_{c}$, where the AF order is suppressed, implying the interplay of a FS change and low energy excitations driven by spin and valence fluctuations. The difference between the TEP signal in the PPM state above H$_{c}$ and in the paramagnetic state (PM) above P$_{c}$ can be explained by different  FS. Band structure calculations at $P=0$ stress that in the AF phase the $4f$ contribution at the Fermi level (E$_F$) is weak while it is the main contribution in the PM domain. By analogy to previous work on CeRu$_2$Si$_2$, in the PPM phase of CeRh$_2$Si$_2$ the $4f$ contribution at $E_F$ will drop. 
\end{abstract}


\maketitle

\section{Introduction}
The interplay between spin fluctuations, valence fluctuations and charge ordering with Fermi surface (FS) instabilities is a key point in the understanding of quantum criticality in itinerant electronic systems. Heavy fermion materials are important examples as their low renormalized characteristic temperature gives the opportunity to switch for example from long range antiferromagnetic (AF) to paramagnetic (PM) ground states under moderate pressure ($P$) of a few GPa and to recover, by applying magnetic field ($H$), a polarized paramagnetic state (PPM) above the AF-PM transition at $H_{c}$ \cite{Lohneysen2007}. An additional particularity is that the magnetically polarized phase at $H>H_{c}$ is often associated with the possibility of a FS reconstruction. This is characteristic of a polarized paramagnetic phase with a FS quite different from the PM ones. Up to now, the CeRu$_{2}$Si$_{2}$ series is the fingerprint example of these phenomena \cite{Flouquet2005a, Aoki2014}. Here we present the effects of $H$ an $P$ on the thermoelectric power (TEP) of the compound CeRh$_{2}$Si$_{2}$ with the aim to observe the response on a system where a FS instability under pressure is well established by previous de Haas van Alphen (dHvA) experiments \cite{Araki2001, Araki2002, Abe1998} and where the magnetic structure of the AF phase has been fully determined \cite{Kawarazaki2000}.\\	
\begin{figure}[h!]
\begin{center}
	\includegraphics[width=8cm]{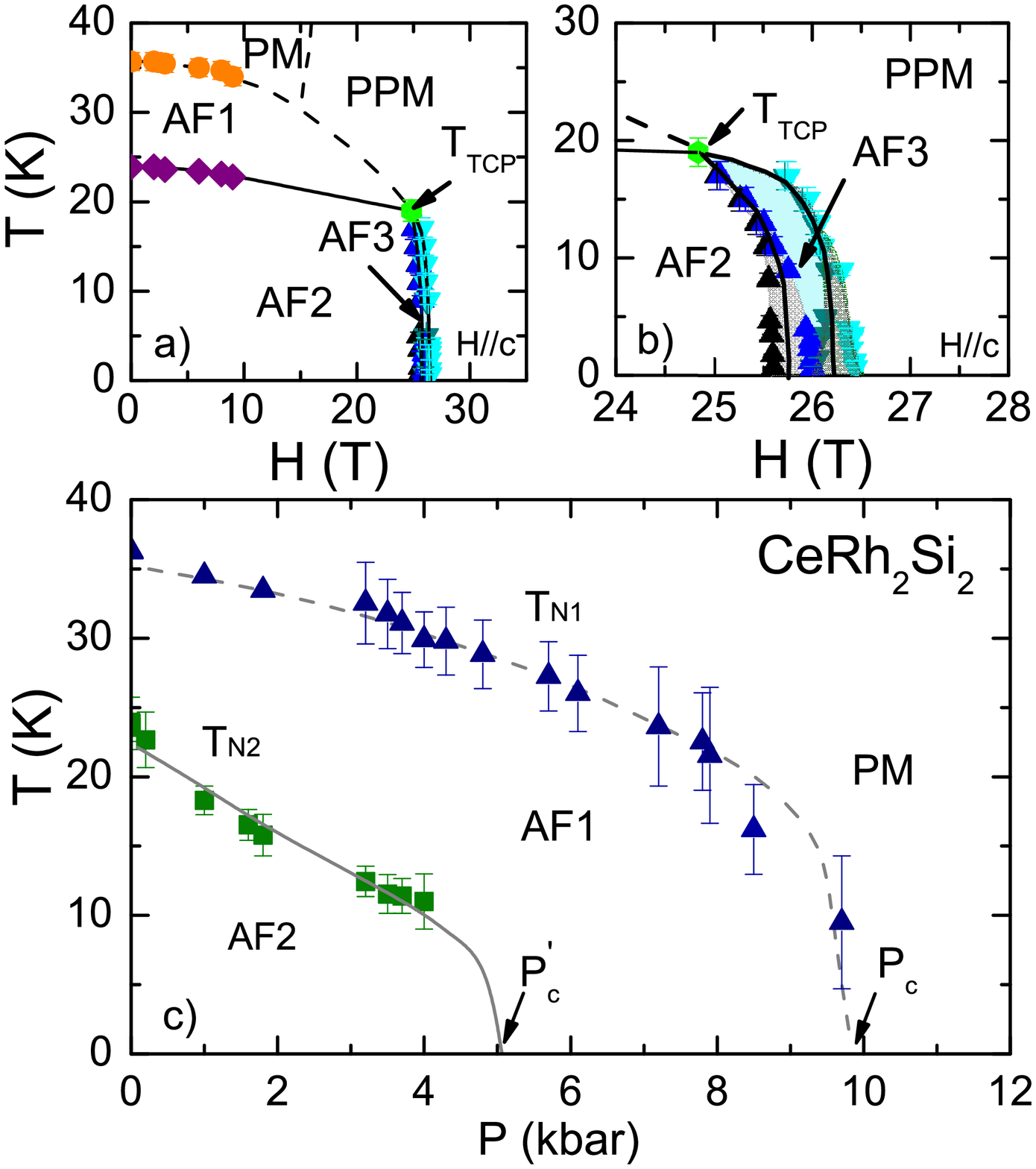}
\end{center}
\caption{\label{Fig1} (Color online) a) $(T, H)$ phase diagram of CeRh$_2$Si$_2$ drawn from field and temperature dependence of TEP for $H \parallel c$. b) Zoom of the $(T, H)$ phase diagram near the suppression of the magnetic order. The light gray and the gray areas indicate the field-hysteresis evolution of the transition from AF2 to AF3 and from AF3 to PPM. The blue areas in these phase diagrams show the narrow $AF3$ phase. c) $(T, P)$ phase diagram of CeRh$_2$Si$_2$ drawn from temperature dependence TEP measurements, $S(T)$, with $J \parallel a$ at different pressures. The triangles show the PM-AF1 transition and the squares the AF1-AF2 one. The error bars correspond to the width of the transition. The suppression of the AF2 and AF1 phase were estimated at $P^{'} _c \sim5$~kbar and $P_c \sim9.8$~kbar, respectively. The dashed and black lines represent second and first order phase transitions, respectively.}
\end{figure}
Figure 1 represents the main features of the CeRh$_{2}$Si$_{2}$ ($H$, $T$, $P$) phase diagram as measured from our TEP. On cooling at $H=0$ and $P=0$, CeRh$_{2}$Si$_{2}$ shows a first AF1 transition at $T_{N1}=36$~K characterized by the wave vector $q_{1}$=($\frac{1}{2}$, $\frac{1}{2}$, 0). The magnetic structure is changing at $T_{N2}$=25~K and in the second AF2 phase, an additional wave vector $q_{2}$=($\frac{1}{2}$, $\frac{1}{2}$, $\frac{1}{2}$) appears\cite{Kawarazaki2000}.
The sublattice magnetization ($M_{0}$) per Ce atoms is near 1.5$\mu_{B}$ and the Sommerfeld coefficient $\gamma=\frac{C}{T}$ extrapolated at $T\rightarrow0$~K is only 23~mJmol$^{-1}$K$^{-2}$. However if the system would not order magnetically, the entropy balance at $T_{N1}$ suggests an extrapolation of $\gamma$ for the PM phase near 400~mJK$^{-2}$mol$^{-1}~$~\cite{Graf1997,Graf1998}.  CeRh$_{2}$Si$_{2}$ belongs to the heavy fermion compounds with the occurrence of strong magnetic correlations by developing a large molecular field $H_{int}$ \cite{Knafo2010} which is at the origin of the drop of the $\gamma$ term of the specific heat. Another interesting feature is that the dominant Ising character with $M_{0}$ aligned along the $c$ axis governs the magnetic structure properties while the Pauli susceptibility in low field is almost isotropic as if there is a decoupling between Ising local moments and the behavior of the itinerant electrons\cite{Settai1997}.

The AF2 phase disappears under pressure at $P'_{c}\sim5$~kbar while the AF1 phase is suppressed at $P_c\sim10$~kbar \cite{Graf1997,Araki2001}. Under magnetic field the AFM is suppressed (for $T<18$~K) by a cascade of first order transitions AF2 $\rightarrow$ AF3  $\rightarrow$ PPM with two critical metamagnetic fields $H_{2-3}$=25.7~T and $H_{c}$=26~T \cite{Settai1997,Abe1997}. As the AF3 phase exists in a narrow magnetic field domain, the main focus will be given later by the change between the AF and PPM phases. Various macroscopic measurements show that at high magnetic field above $H_{c}$, the PPM ground state differs from the PM ones \cite{Knafo2010}. Up to now, high magnetic field FS studies have been restricted to the AF phase, i.e to $H < H_{c}$, and FS measurements have been mainly performed under pressure with $H \parallel c-axis$ in the basal plane \cite{Araki2001}. Under pressure FS reconstructions have been observed at $P'_{c}$ and $P_{c}$ \cite{Kawarazaki2000,Movshovich1996,Boursier2005}. A rough interpretation suggests a localized description of the $4f$ electrons below $P_{c}$ and an itinerant description above $P_{c}$ \cite{Araki2001}.
\section{Experimental details}
High quality single crystals of CeRh$_2$Si$_2$ were grown using the Czochkralski method in a tetra-arc furnace. Single crystal ingots were oriented by Laue photograph and were cut by a spark cutter. No heat treatment was done for the present samples. The sample quality was checked by resistivity measurements. The residual resistivity ratio (RRR) is in the range from 30 to 100. The highest quality sample was used for quantum oscillations measurements. 

The TEP study was performed with $H\parallel c$ in order to reach the PPM regime at $P=0$. The experiment was performed in longitudinal (thermal gradient $\nabla T\parallel c$ axis) and transversal configurations ($\nabla T\parallel a$ axis). $H$ scans were done with a conventional superconducting magnet ($H<16$~T) and with a resistive magnet at the LNCMI laboratory ($H<28$~T). Zero magnetic field measurements under pressure were performed in a piston cylinder cell. Details of the experimental realization of the high pressure experiment can be found in Ref.~\onlinecite{Palacio}.

\begin{figure}[h!]
	\begin{center}
	\includegraphics[width=8cm]{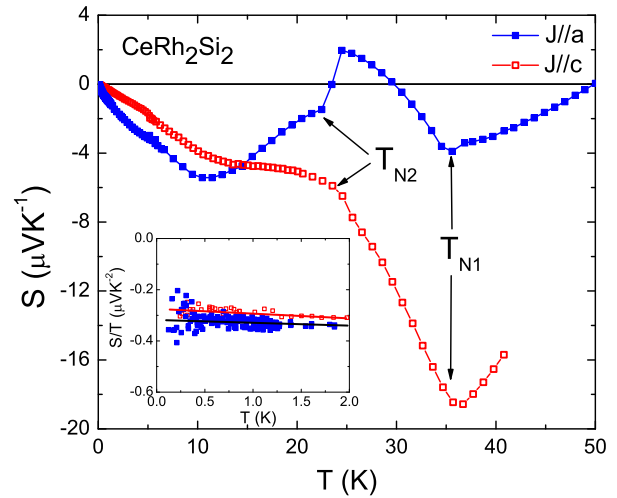}
	\end{center}
	\caption{\label{Fig2} (Color online) Temperature dependence of the TEP $S(T)$ for $J\parallel a$ filled symbols and $J\parallel c$ open symbols. The inset shows that $S(T)/T$ is constant below 2~K for both configurations. The straight red and black lines of the inset indicate the extrapolation of $S(T)/T$ at $T \rightarrow 0$ for $J\parallel c$ and $J\parallel a$ configurations, respectively.}
\end{figure}

\section{Results}
Figure \ref{Fig2} shows the temperature dependance of TEP at ambiant pressure for both, $J\parallel a$ and $J\parallel c$ configuration, at $H=0$. In both configurations the TEP shows abrupt changes in the $T$ dependence at the magnetic transitions. At very low temperature below $T_{min}\sim10$~K, the TEP is mainly isotropic while a strong anisotropy is detected above $T_{min}$. The inset of Fig.~\ref{Fig2} shows the temperature variation of $S/T$ below $T=2$~K. In agreement with the small value of the Sommerfeld coefficient $\gamma$, $S/T$ is rather weak for an heavy fermion compound. Furthermore, the usual positive value of $S/T$ as $T\rightarrow0$~K observed for many Ce based heavy fermion compounds is not recovered. Clearly, the creation of new small Brillouin zones and its associated FS reconstruction modify the band structure and thus the energy derivative of the density of states which is directly linked to the TEP signal. As discussed below the Ce-$4f$ contribution to the density of states at the Fermi level is weak in the AF phase.

\begin{figure}[h!]
	\begin{center}
	\includegraphics[width=8.5cm]{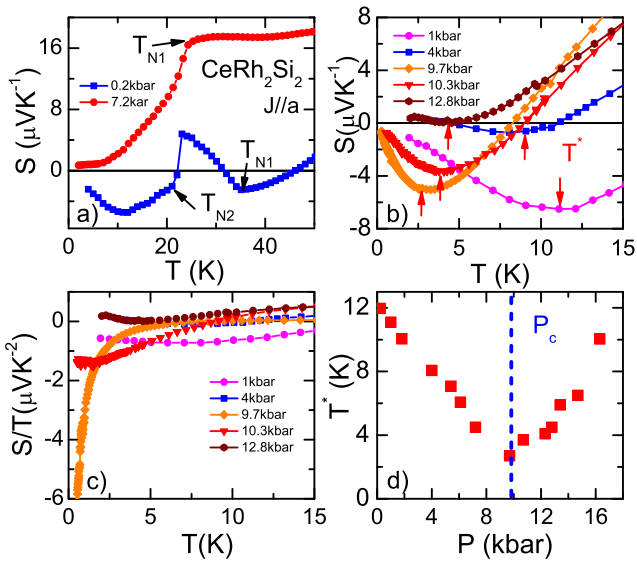}
	\end{center}
	\caption{\label{Fig3} (Color online) a) Temperature  dependence of the TEP, $S(T)$, for $J \parallel a$ at $P$=0.2~kbar and $P$=7.2~kbar for CeRh$_2$Si$_2$. b) $S(T)$ for different pressures at low temperature for $J \parallel a$. The temperature positions, $T^*$, of the minimum/inflection points of $S(T)$ are indicated by red vertical arrows. c) $S(T)/T$ at low temperature for different pressures. The extremely large value of $S(T)/T$ for $T \rightarrow 0$ at $P$=9.7~kbar reflects the proximity to the critical pressure $P_{c}$. d)As a function of pressure, the strong decrease of $T^*$, which generally marks the entrance in a coherent low temperature regime, is a manifestation of critical fluctuations near $P_{c}$.}
\end{figure}

A strong variation of the TEP under pressure is observed notably close to $P'_{c}$ and $P_{c}$. Typical $S(T, P)$ curves are presented in Fig.~\ref{Fig3}a). From the marked signature of $S(T)$ at $T_{N1}$ and $T_{N2}$, the ($T, P$) phase diagram of CeRh$_{2}$Si$_{2}$ was drawn in Fig.~\ref{Fig1} in good agreement with previous measurements \cite{Araki2002,Kawarazaki2000,Movshovich1996,Boursier2005}. $S(T)$ and $S(T)/T$ at low temperature for different pressures are represented in figure \ref{Fig3}b) and c) respectively. The extremely large value of $S(T)/T$ for $T\rightarrow 0$ for $P$=9.7~kbar reflects the proximity to the critical pressure $P_{c}$. The entrance in a coherent regime is marked initially by a well defined minimum at $T^*$. In Fig.~\ref{Fig3}d) the strong decrease of $T^*$(extracted from $S(T)$ in Fig.~\ref{Fig3}b)) is a manifestation of critical fluctuations near $P_{c}$. 

\begin{figure}[h!]
	\begin{center}
	\includegraphics[width=8cm]{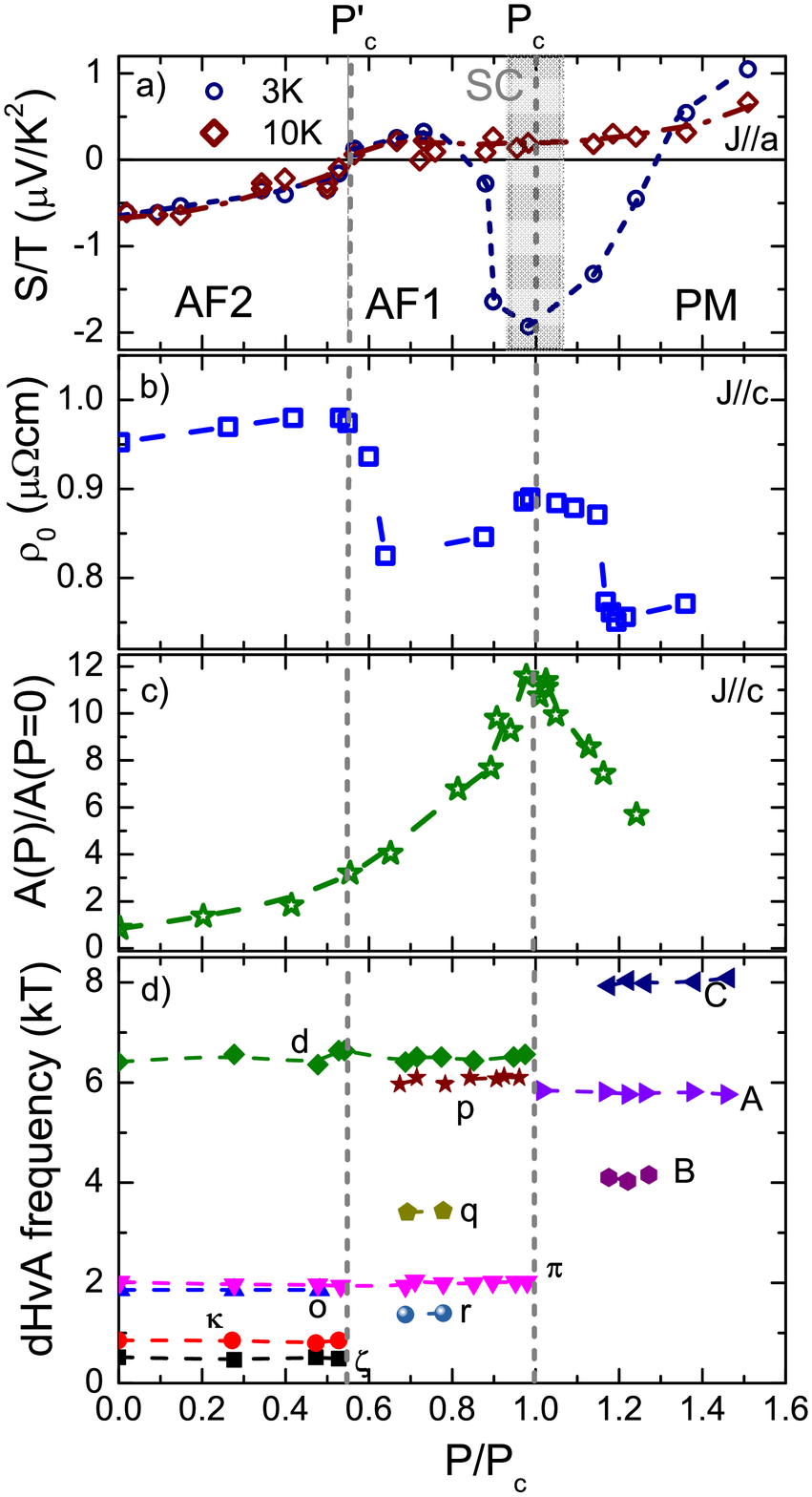}
	\end{center}
	\caption{\label{Fig4} (Color online) Observation of FS changes under pressure by different probes: a) TEP over temperature,$S/T$, at $T=3$~K (open blue circles) and at $T=10$K (red diamond), b) residual resistivity $\rho _0$ (open blue squares), c) normalized $A$ coefficient $A(P)/A(P=0)$ (green open stars)\cite{Araki2002} and d) dHvA frequencies (full symbols)\cite{Araki2001} as a function of pressure $P/P_c$.}
\end{figure}

The pressure dependances of the TEP  at 10~K and 3~K are illustrated in Fig.~\ref{Fig4}a). Below 0.8~kbar, rather similar values of $S/T$ are obtained, however in the critical regime (8-13~kbar) large differences exist between $T=10$~K and $T=3$~K  in agreement with the occurrence of a deep negative minimum on cooling close to $P_{c}$. In this pressure range, the characteristic electronic energy drops on approaching $P_{c}$ as clearly observed in the decrease of $T^*$ (Fig.~\ref{Fig3}d)) and in the shrinking of the low temperature domain (at $P_{c}$ lower than 8~K) where the $AT^2$ Fermi liquid resistivity is detected \cite{Boursier2005}.

Furthermore in Fig.~\ref{Fig4}, different anomalies detected are recorded in the $P$ dependence of different quantities: b) the residual resistivity $\rho_{0}$, c) the $A$ coefficient \cite{Araki2002} and d) the dHvA frequencies \cite{Araki2001}. The signatures of the TEP at $P'_{c}$ and  $P_{c}$ corresponds to clear changes in the residual resistivity and dHvA frequencies. The inelastic $T^2$ term of the resistivity has a marked maximum while in Ref.~\onlinecite{Boursier2005} a plateau has been reported. Further experiments must solve the origin in the resistivity anomalies detected near $P_{c}$ for  $J_{e}\parallel a$ and $J_{e}\parallel c$ \cite{Araki2002}.
\begin{figure}[h!]
	\begin{center}
	\includegraphics[width=8cm]{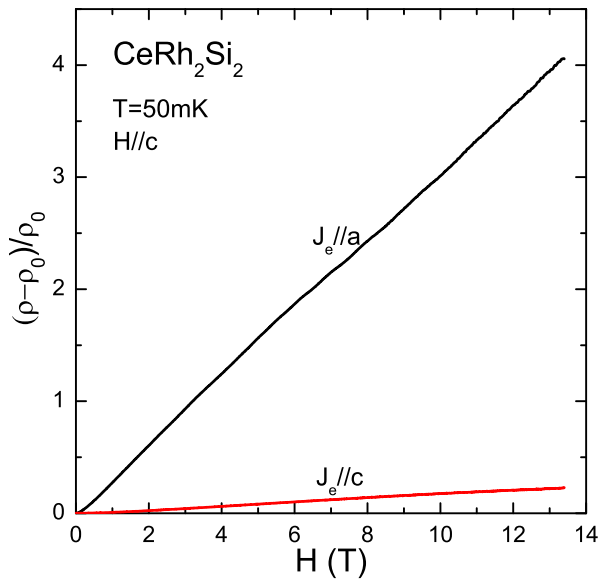}
	\end{center}
	\caption{\label{Fig5} (Color online) ($\rho -\rho _0)/\rho_0$ as a function of the magnetic field for transverse ($RRR=100$) and longitudinal ($RRR=30$) configurations. A strong magnetoresistance increase is observed in the high quality sample.}
\end{figure}
This shows that electronic fluctuations (mainly spin fluctuations) are detected far below $P_{c}$. The important observation in the TEP is the switch of $S/T$ from a negative to a positive value at $P'_{c}$, the deep negative minimum at $P_{c}$ and the recovery of the common positive sign of the TEP for Ce heavy fermion compounds above $P_{c}$. The drastic change of $S$ around $P_{c}$ is clearly associated with the large FS reconstruction detected at $P_{c}$ (Fig.~\ref{Fig4} d)).  At least the rather large pressure width of the $S(P)$ anomaly ($\Delta p\sim0.3$~GPa) coincides with the enhancement of the $A$ coefficient. CeRh$_{2}$Si$_{2}$ is thus a key example of the consequence of the interplay between electronic fluctuations (as we will discuss spin and valence fluctuations) at $P_c$ and a FS instability.
\begin{figure}[h!]
	\begin{center}
	\includegraphics[width=8cm]{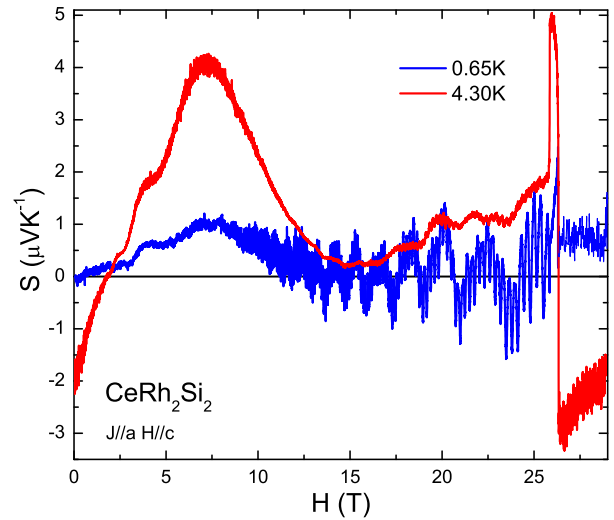}
	\end{center}
	\caption{\label{Fig6} (Color online) Isothermal TEP measurements as a function of magnetic field $S(H)|_T$ at different temperatures for the transverse configu
	ration. The low temperature curve shows the suppression of quantum oscillations at the PPM phase. The higher harmonic content of the Seebeck oscillations imparts a characteristic sawtooth shape to the oscillations.}
\end{figure}
 \begin{figure}[h!]
	\begin{center}
	\includegraphics[width=8cm]{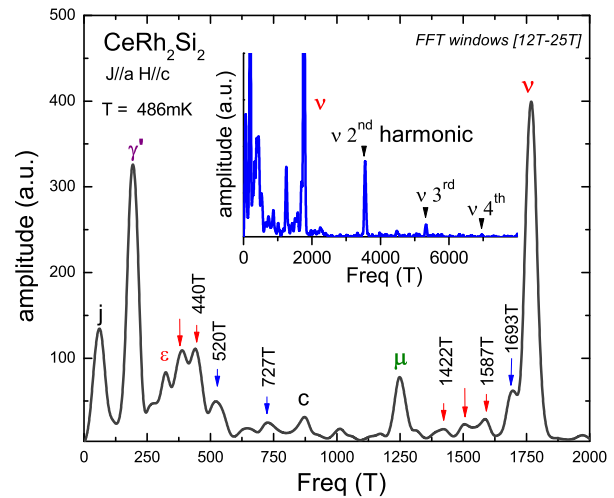}
	\end{center}
	\caption{\label{Fig7} (Color online) FFT spectrum of the AF2 phase of CeRh$_2$Si$_2$ obtained from isothermal TEP measurement at the lowest temperature $T=486$~mK with a field range from 12~T to 25~T. The known-frequencies are named following Ref.~\onlinecite{Araki2001}, the new ones are indicated by blue arrows and the frequencies due to magnetic breakdown by red arrows. The inset shows the main frequency $\nu$ and its harmonics up to the fourth one.}
\end{figure}

Let us now focus on the field dependance of TEP at $P=0$. As in this multi-band heavy fermion system the signal of one subband is weighted by their respective electrical conductivity \cite{Barnard}, large magnetoresistivity may reveal different signatures. In our case, for the transversal configuration, an excellent crystal was used with RRR$\sim100$ while for the longitudinal configuration a crystal with only RRR$\sim30$ was measured. Figure \ref{Fig5} emphasizes the large magnetoresistivity detected for the $J_e~\parallel a$ by comparison to the $J_e \parallel~c$ ones. 

The consequence in TEP is that, for the transverse configuration, features associated with the entrance in the collision-less regime were observed already for $T \sim6$~K, as shown by the observation of large quantum oscillations in the TEP for the transverse configuration up to 4.3~K (see Fig.~\ref{Fig6}). Surprisingly, already at $T=4.3$~K for $H\sim5$~T quantum oscillations are detected up to $H_{c}\sim26$~T while no quantum oscillations signal is detected above $H_{c}$. To get information of the FS properties of CeRh$_{2}$Si$_{2}$, we performed a Fast Fourier Transform (FFT) of the Seebeck signal. The spectrum for the AF2 phase of CeRh$_{2}$Si$_{2}$ for the highest quality sample (transverse configuration) is represented on figure \ref{Fig7}. No oscillations have been observed in the spectrum for the AF3 and PPM phases above $H_{c}$. The higher harmonic content of the TEP oscillations imparts a characteristic sawtooth shape to the oscillations up to the $4^{th}$ harmonics of $\nu$ has been observed. The frequencies in the TEP signal and those reported previously from dHvA experiments \cite{Araki2001} are listed in table 1. The frequencies detected in TEP are in good agreement with those from dHvA, although TEP measurements cannot detect the low frequencies of the FS of the AF2 state. Similarly, the various masses extracted from the TEP branches \cite{Palacio} are similar to the ones obtained in the dHvA experiments \cite{Araki2001}: the measured effective masses go from 0.64~m$_{0}$ for the $\gamma$ branch to 5~m$_{0}$ for the $\mu$ branch \cite{Araki2001,Abe1998}. However, TEP additional frequencies have been detected corresponding to orbits of the AF2 FS which have been never detected except for frequencies close to 387~T and 1505~T which correspond to a magnetic breakdown \cite{Palacio}. Careful analysis of the FS determination in relation with new band structure calculation will be presented separately. Experimental studies of the TEP quantum oscillations can be found in Ref.~\onlinecite{Palacio}.
\begin{table}[h]
\begin{ruledtabular}
\caption{\label{Table}List of quantum oscillation frequencies in the $AF2$ phase obtained from dHvA measurements of Ref.~\onlinecite{Araki2001} and from $S(H)$ at $T=468$~mK for the transverse configuration.\\}
\centering
\begin{tabular}{ccc}
  \multicolumn{2}{c}{Frequencies of the $AF2$ phase (T)}\\
  \hline
  branches & dHvA (Ref.~\onlinecite{Araki2001}) & $S(H)$\\
   & $[13-16.9]$~T & $[12-25]$~T\\
  \hline
  l& $44$ &  \\
  k& $56$ &  \\   
  j& $66$ & $61$ \\
  & $77$ &  \\ 
  $\alpha ^{''}$& $81$ &  \\
  $\gamma'$& $137$ &  \\
  $\gamma $& $184$ & $194$ \\
  $\varepsilon$& $327$ & $324$ \\
   & & $387$ \\
   & & $440$ \\
   & & $520$ \\
   & & $727$ \\
  c& $804$ & $870$ \\
  $\mu$& $1160$ & $1249$ \\
   & & $1422$ \\
   & & $1505$ \\
   & & $1587$ \\
   & & $1693$ \\
  $\nu$& $1770$ & $1768$ \\
  a & $7560$ &  \\
\end{tabular}
\end{ruledtabular}
\end{table}

\begin{figure}[h!]
	\begin{center}
	\includegraphics[width=8cm]{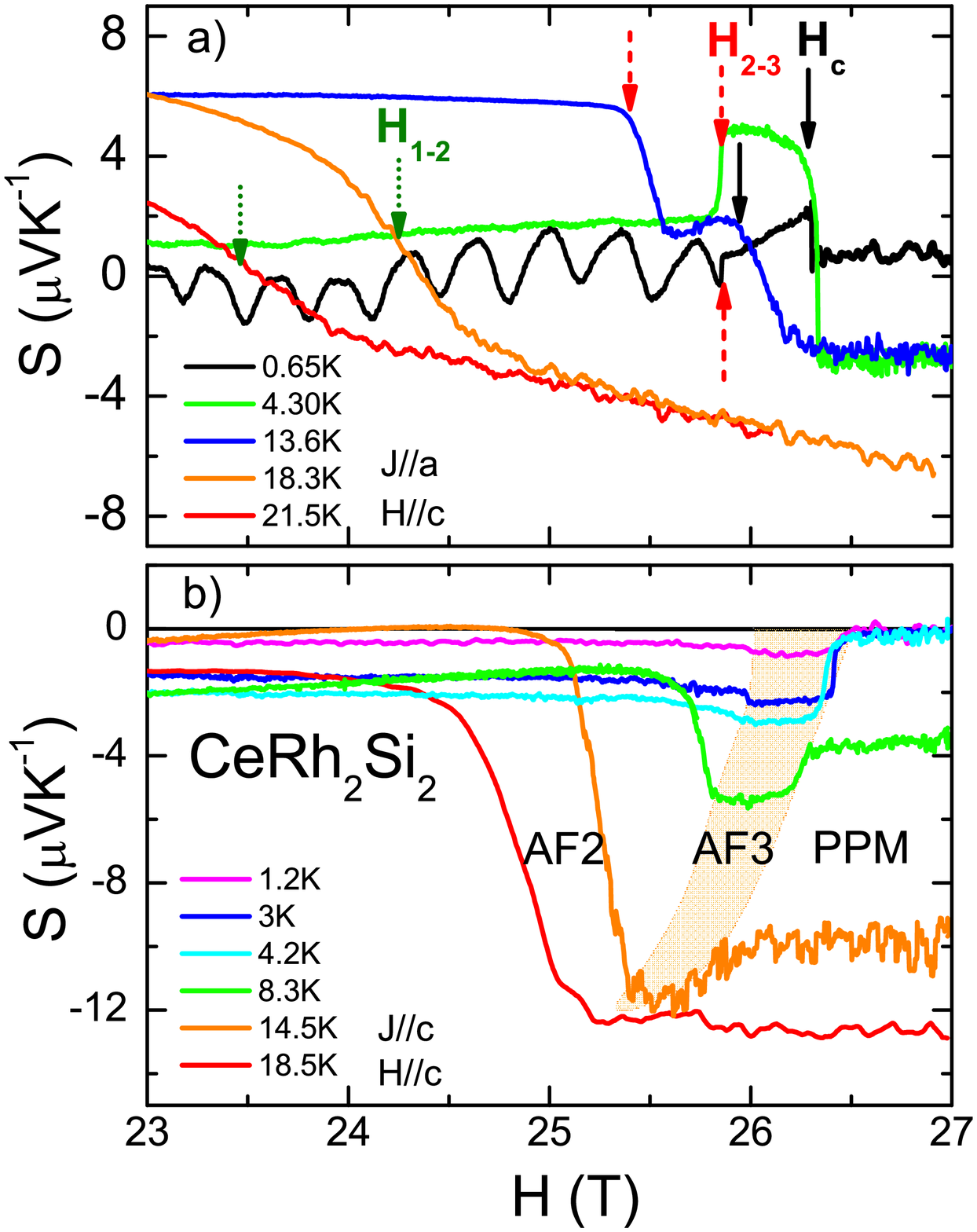}
	\end{center}
	\caption{\label{Fig8} (Color online) Isothermal TEP, $S(H)$, at different temperatures focusing on the high field range. a) For $J\parallel a$, the metamagnetic transitions $H_{2-3}$ and $H_c$ are characterized by a two step anomaly and symbolized by red dashed and black vertical arrows, respectively. The AF1-PPM transition ($H_{1-2}$) which takes place for $T>T_{TCP}$=17~K, is marked by green dotted vertical arrow. b) For $J\parallel c$, the evolution of the AF3, delimited by the orange area, is represented.}
\end{figure}

As shown in Fig.~\ref{Fig8},  the transitions AF2-AF3 and AF3-PPM are characterized by a rapid change in the absolute value of the TEP coefficient and by an hysteric behavior confirming the first order nature of the transitions. Moreover, these transitions show a change of TEP  in the low temperature regime. At the lowest temperature ($T=0.65$~K), $S(H)$ shows large quantum oscillations up to the AF2-AF3 transition at $H_{2-3}$=25.7~T. At the AF3-PPM transition at $H_c=26$~T a very sharp drop of $S(H)$ appears. A higher temperature ($T=13.6$~K), $S(T)$ drops sharply at $H_{2-3}$ and $H_c$. For $T>T_{TCP}$=17~K, the transition, AF1-PPM, becomes broad  with no hysteresis and the broadening marks the suppression of the AF order with a second order transition. For the longitudinal configuration, see Fig.~\ref{Fig8} b), the evolution of the AF3 phase, can also be observed. The collapse of the AF3 phase occurs at $T\sim T_{TCP}$. The ($H$, $T$) phase diagrams determined with these different anomalies (with $\nabla T\parallel$ c and  $\nabla T\parallel a$) are in excellent agreement as can be seen in Fig.\ref{Fig1}. \cite{Knafo2010,Settai1997,Abe1997}, the slight difference in the position of the AF2  boundary is only due to the difficulty to select precisely the $S(T)$ anomalies above $T_{TCP}$.

\section{Discussion}
\begin{figure}[h!]
	\begin{center}
	\includegraphics[width=8cm]{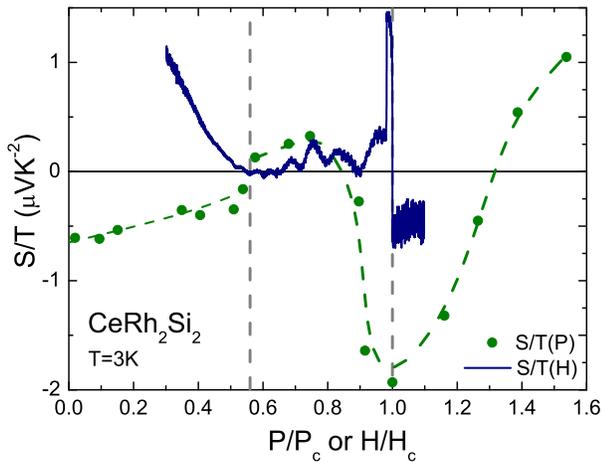}
	\end{center}
	\caption{\label{Fig9} (Color online) $S/T$ for the transverse configuration as a function of magnetic field $H/H_c$ and as a function of pressure $P/P_c$ in blue continuous line and green dots, respectively.}
\end{figure}

Drastic changes of the TEP are observed on crossing $H_{c}$ and $P_{c}$. Figure \ref{Fig9} shows at $T= 3$~K the respective $P$ and $H$ dependence of $S$ as function of the reduced parameters $\frac{P}{P_{c}}$ and $\frac{H}{H_{c}}$. The large width of the $P$ anomaly across $P_{c}$ contrasts with the sharp feature of the field scan at $H_{c}$. In magnetic field scans, two main features occur: 
i) the disappearance of the quantum oscillations above $H_c$ pointing out major FS change connected with the unfolding of the Brillouin zone,
ii) a step in the macroscopic signal reflecting clearly the strong metamagnetic transition at $H_{c}$ (magnetization jump $\Delta M\sim1.2$~$\mu_B$ per Ce atom) coupled with a strong change of dominant heat carriers.
Under pressure, the quantum phase transition at $P_{c}$ is marked by a broad negative signature directly linked with anomalies in the residual resistivity, in the Hall effect \cite{Boursier2005}, and in the $P$ dependance of the sublattice magnetization $M_{0}$  which drastically collapses. Evidence is given from thermal expansion measurements \cite{Villaume2008} that the quantum phase transition at $P_{c}$ is weakly first order. Also the $T$ domain of the $AT^2$ dependence of the resistivity does not shrink to zero as predicted and observed for AF second order quantum critical point\cite{Lohneysen2007}: a $T^2$ dependence of $\rho$ is observed for all pressures \cite{Araki2002,Boursier2005}.    

  \begin{figure}[h!]
	\begin{center}
	\includegraphics[width=8cm]{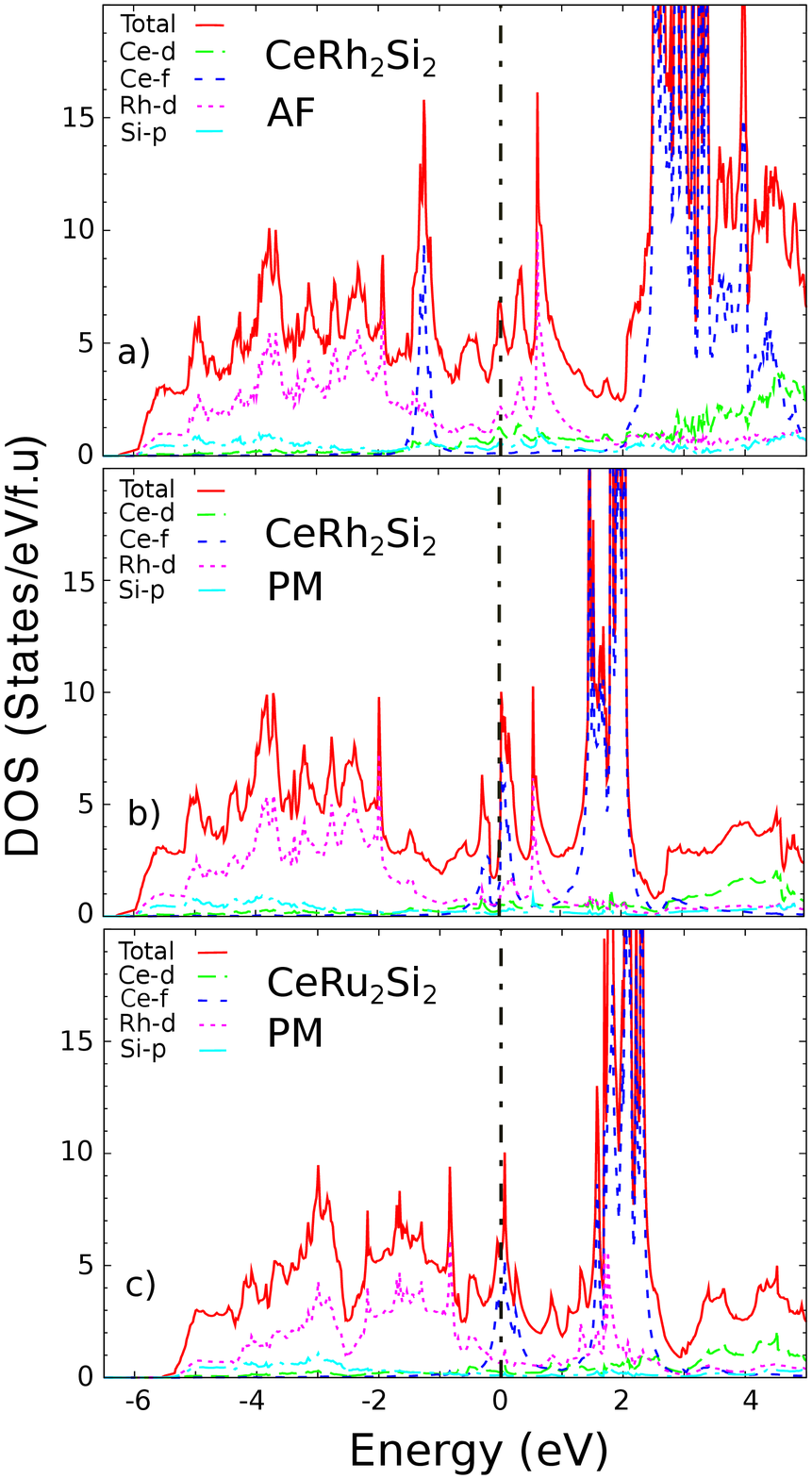}
	\end{center}
	\caption{\label{Fig10} (Color online). Comparaison between the densities of states determined by LDA+U calculation in CeRh$_2$Si$_2$ and CeRu$_2$Si$_2$: a) and b) corresponds to the 4q-AF and PM phases of CeRh$_2$Si$_2$, respectively and c) to the PM state of CeRu$_2$Si$_2$. In the 4q-AF state of CeRh$_2$Si$_2$, no Ce-$4f$ contribution appears at the FS in contrary, for the PM state, $f$-states are now located around the Fermi level and have a large contribution to the FS. The densities of states in the PM state of CeRu$_2$Si$_2$ and CeRh$_2$Si$_2$ are very similar.}
\end{figure}
The first order nature of the transition at $P_{c}$ appears in good agreement with the drastic change of the FS reported in dHvA experiments \cite{Araki2001}. Far below $P_{c}$, at ambient pressure, the $4f$ electrons are considered to be localized as no main $4f$ contribution appears at the Fermi level. De facto the angular dependence of the large dHvA frequencies can be explained by those of LaRh$_2$Si$_2$. Due to the magnetic order, the magnetic Brillouin zone is 8 times smaller than that of body centered tetragonal Brillouin zone. The density of states, calculated by an LDA+U method with the U=5eV in the framework of full potential linearized augmented plane wave (FLAPW) method are represented on Fig.~\ref{Fig10}. We performed the calculations with self-consistently determining the charge density and the density matrices for the nonmagnetic state and with assuming Ce-f$^1$ occupation for the $|j=5/2, j_z=5/2>$ orbital of the density matrices for the magnetic state, taking the eight CeRh$_2$Si$_2$ formula magnetic unit cell for the 4q-magnetic order with the proper magnetic alignment\cite{Kawarazaki2000}. For the 4q-AF state, there is no direct contribution from the Ce-$4f$ states at the FS but in the PM state the $4f$ states are now located around the Fermi level and have large contribution to the FS. This new approach opens the possibility to understand the FS topology with an initial itinerant picture by incorporating the large molecular field ($H_{int}$) created by the robust AF (concomitant large $T_{N}$ and $M_{0}$) and the creation of a new Brillouin zone. Let us point out that the PM density of states of CeRh$_2$Si$_2$ is quite similar to the one obtained for CeRu$_2$Si$_2$; the extension of the calculation for CeRu$_2$Si$_2$ to strong magnetic polarization shows that in CeRu$_2$Si$_2$ a deep shift of the $4f$ level from the FS occurs above its metamagnetic transition\cite{Suzuki2010}.\\
 \begin{figure}[h!]
	\begin{center}
	\includegraphics[width=8cm]{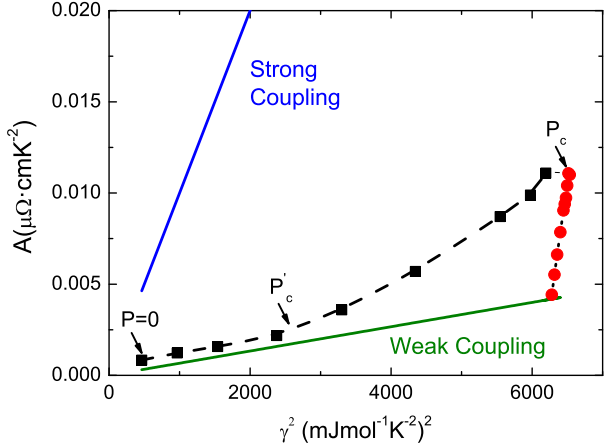}
	\end{center}
	\caption{\label{Fig11} (Color online) $A$ coefficient (values obtained from Ref.~\onlinecite{Boursier2005}) as a function of the square of the electronic specific heat $\gamma ^{2}$ (from Ref.~\onlinecite{Graf1997}). The blue and green lines represent the limits for strong coupling and weak coupling behaviors. The black square symbols correspond to he AF phase and the red circles to the PM phase.}
\end{figure}
It is appealing to propose that the sign change of $S/T$ at 3~K right at $P'_{c}$ marks the switch from a localized to an itinerant nature of the $4f$ electrons as it was proposed for a Kondo lattice\cite{Hoshino2013} inside the AF domain with the particularity of very low energy strong local fluctuations driven by a Lifshitz transition. Experimental evidences for such a scenario have been reported for the system CeRu$_{2}$(Si$_x$Ge$_{1-x}$)$_{2}$ \cite{Aoki2014}. However, as shown in Fig.~\ref{Fig4}d) no drastic change of dHvA frequencies has been observed and the $d$ and $\pi$ branches are almost not affected at $P'_c$. The $\kappa$ branch has not been observed above $P'_c$. These small differences in the observed branches below and above $P'_c$ seem to reflect the Brillouin zone change between the AF1 and AF2 phases due to the different ordering vectors. Furthermore, there is no signature of energy decrease of the excitations in TEP.  As the transition AF2-AF1 is of first order at $P'_{c}$ \cite{Araki2001,Knafo2010}, the possibility of additional electronic Lifshitz transition at  $P'_{c}$ remains. There is strong deviation from $AT^2$ resistivity behavior at $P'_{c}$ but it can be explained in the frame of AF2-AF1 inhomogeneous mixing\cite{Boursier2005}. In addition, there is no doubt that above  $P_{c}$ the new FS is in agreement with $4f$ itinerant quasiparticles.  
  \begin{figure}[h!]
	\begin{center}
	\includegraphics[width=8cm]{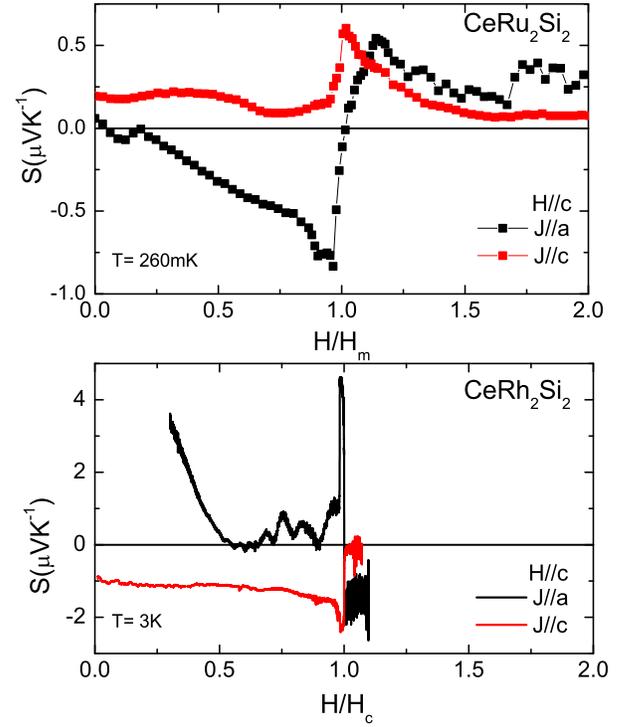}
	\end{center}
	\caption{\label{Fig12} (Color online) Comparaison between the isothermal TEP measurements as a function of magnetic field of CeRu$_2$Si$_2$ and CeRh$_2$Si$_2$ for transverse (red markers) and longitudinal(black markers) configurations. The magnetic field scales of this two compounds are normalized by the critical magnetic field $H_m$ and $H_c$ for CeRu$_2$Si$_2$ and CeRh$_2$Si$_2$, respectively. A strong modification of the isothermal thermopower measurement has been observed at the transition in both $Ce$ compounds.}
\end{figure}

At $P_{c}$, CeRh$_2$Si$_2$ is very close to enter in the mixed valence regime, i.e $P_{c}$ close to $P_{v}$ the characteristic valence pressure \cite{Flouquet2005a,Watanabe2013}. Proofs of this proximity are given (i) by the $P$ decrease of the magnetic anisotropy (the maximum of the susceptibility ratio between c and a axis varying from 6 at $P=0$ to already 2 at 0.098~GPa) \cite{Mori1999}, (ii) by the strong $P$ variation of the thermal expansion ratio $\frac{\alpha_a}{\alpha_c}$ (going from 0.23, 0.73 and 1 for $P$=0.041, 0.077 and 0.104~GPa) \cite{Villaume2008}, (iii) by the weak value of the $\frac{A}{\gamma^2}$ Kadowaki Woods (KW) ratio on both sides of $P_{c}$. The KW ratio is drawn in  Fig.\ref{Fig11} taking the $A$ value from Ref.~\onlinecite{Araki2002} and $\gamma$ from Ref.~\onlinecite{Graf1997}, respectively. In the case of strong coupling (Ce level degeneracy N=2), $\frac{A}{\gamma^2}$ is predicted to reach $10^{-5}$~$\mu \Omega$cm(Kmole/mJ)$^2$ while in the case of weak coupling (N=6 full Ce degeneracy) the ratio is reduced by an order of magnitude ( $\frac{A}{\gamma^2}\sim\frac{10^{-5}}{N(N-1)}$ )\cite{Miyake1989,Tsujii2005}. Due to the large strength of H$_{int}$ for $P \ll P_{c}$ and the low contribution of the $4f$ localized component at the Fermi level, $\frac{A}{\gamma^2}$ is near the weak coupling limit. Strong increase of KW ratio is observed on approaching $P_{c}$, but on increasing slightly the pressure above $P_c$, $\frac{A}{\gamma^2}$ tends again towards the weak coupling case. Let us just note that for the case of YbRh$_2$Si$_2$ (discussed later) a strong coupling is observed \cite{Gegenwart2008}. The main difference in CeRh$_2$Si$_2$ is that the interplay of rather comparable Kondo and AF correlations restores a local Ce picture with well resolved crystal field levels \cite{Willers2012}. But on increasing pressure, assuming a $P$ invariance of the bare crystal field (with characteristic energy $k_BT_{CF}$), the concomitant $P$ increase of the Kondo energy and $P$ collapse of AF interactions leads to quench the full angular momentum via the Kondo coupling: the crystal field is no more efficient to operate. Nevertheless, further high pressure measurements, far above $P_{c}$, must be performed to fully validate this hypothesis.

It is worthwhile to compare the TEP signal of CeRh$_2$Si$_2$ with the one of CeRu$_2$Si$_2$ as this system, already in the PM state at $P=0$ ($P_{c}\sim-0.4$~GPa), is a reference in the study of pseudometamagnetic phenomena.  Pseudomagnetism in CeRh$_2$Si$_2$ appears above $P_{c}$ for a magnetic field $H_{m}$ ($H_{m}=$40~T close to P$_{c}$ \cite{Hamamoto2000}). $H_{m}$ can be interpreted as a crossover continuation of the metamagnetic field $H_{c}$ \cite{Flouquet2005a,Amato1989,Pourret2014}. Expanding the volume via La or Ge dopings leads to recover AF long range order and first order metamagnetism under magnetic field \cite{Flouquet2005a, Aoki2014}. We will focus here on the difference in magnetic response of the undoped pure lattice.
At ambient pressure at the metamagnetic field $H_{c}\sim26$~T, the Sommerfeld coefficient of CeRh$_2$Si$_2$ reaches 80~mJmole$^{-1}$K$^{-2}$ while CeRu$_2$Si$_2$ at $H_{m}$ reaches a value of $\gamma\sim600$~mJmole$^{-1}$K$^{-2}$. We notice that the factor 10 between the $\gamma$ values is related to the ratio of their relevant energies, then we can compare their TEP response taking into account this factor 10 in temperature analysis.. Figure \ref{Fig12} shows the longitudinal and transverse TEP responses of CeRu$_2$Si$_2$ at 290~mK by comparison to the CeRh$_2$Si$_2$ ones at 3~K. At $H_{m}$ in CeRu$_2$Si$_2$ a drastic change of the FS has been reported \cite{Aoki2014,Aoki1995, Julian1994} and is associated with a strong increase of $A$ and $\gamma$ right at $H_{m}$ preserving a quasi constant $\frac{A}{\gamma^2}$ ratio. Due to the large magnetoresistivity anisotropy, the transverse and longitudinal signals have opposite sign response at $H_{m}$. However, the metamagnetic transitions present quite similar width, $\Delta H\sim$~2T, for transverse and longitudinal configurations. This width is quite larger than the width detected for CeRh$_2$Si$_2$ at $H_{c}$ ($\frac{\Delta H}{H_{c}}\sim$ 0.02). Comparing to UCoAl that presents a first order metamagnetic transition from PM to FM, we notice that a sharp change of $S$ at the transition also occurs in this compound $\cite{Palacio2013}$. Thus, the TEP signal in CeRh$_2$Si$_2$ at $H_c$ is clearly governed by the strong first order nature of the metamagnetic transition while the FS crossover detected at $H_{m}$ in CeRu$_2$Si$_2$ is characteristic of a sole Lifshitz transition.

These new CeRh$_2$Si$_2$ data lead to emphasize its difference with the isostructural Yb compound YbRh$_2$Si$_2$. Previous discussions can be found in refs. \onlinecite{Knebel2006,Boursier2008}.
In case of CeRh$_2$Si$_2$ there is no divergence of $A$ or $\gamma$ coefficient at $P_{c}$ or $H_{c}$. Such phenomena have been reported for the Yb heavy fermion system YbRh$_2$Si$_2$ (T$_{N}\sim70$~mK) as the magnetic field is swept through the AF phase transition ($H_{c}\sim0.06$~T along the easy axis)\cite{Gegenwart2008,Si2001,Steglich2014}. Kondo breakdown has been involved in this system with the new idea of local quantum criticality with fluctuations between small and large FS. Although a signature of a FS change has been observed in Hall effect \cite{Paschen2004} no corresponding variation in the TEP can be detected on approaching $H_{c}=0.6~T$ for $H\parallel c$\cite{Machida2012}. However a Lifshitz transition appears at $H_0\sim10$~T$\gg H_{c} $ as the magnetic polarization reaches a critical value \cite{Pfau2013,Pourret2013b,Boukahil2014}. The common feature of the anomaly at $H_0$ in YbRh$_2$Si$_2$  and here at the first order transition of CeRh$_2$Si$_2$ at $H_c$ is the crossing through a critical value of electronic polarization. Indeed, this mechanism may imply metamagnetism (case of CeRu$_2$Si$_2$) or not (case of YbRh$_2$Si$_2$). 
 
Recently the observation of the itinerant nature of $4f$ electrons in YbRh$_2$Si$_2$ in the PM phase above $T_{N}$ at zero magnetic field \cite{Guttler2014} (in good agreement with previous dHvA experiments \cite{Rourke2008a,Knebel2006}) leads to a strong suspicion on the switch of the $4f$ localization to describe the quantum phase transition. A new approach is so called strong coupling theory of heavy fermion criticality \cite{Abrahams2014}, the basic phenomena being that critical spin fluctuations of the AF wave vector $Q$ induces fluctuations at small $q$ with the consequence of diverging effective mass over the whole FS.

In the weak coupling case of CeRh$_2$Si$_2$, the switch from itinerant to localized picture is driven by the onset of the AF order and its associated Brillouin zone change. The consequence on YbRh$_2$Si$_2$ is still an open question as the magnetic structure has not been determined yet. CeRh$_2$Si$_2$ presents the additional property to lose its magnetic anisotropy under pressure presumably on entering in the mixed valence regime while in YbRh$_2$Si$_2$ the change of the anisotropy on entering in the trivalent regime seems to occur only above 90~kbar \cite{Knebel2006}. As emphasized at $P=0$ in CeRh$_2$Si$_2$ the intersite magnetic coupling ($H_{int}$)  is higher or comparable to the Kondo energy. In YbRh$_2$Si$_2$ it is quite the opposite however the trivalent configuration can be stabilized enough to present very low energy excitation and at the end in the mK range to order magnetically since the 13 electrons of the $4f$ shell ($4f^{13}$) are strongly localized by comparison to the $4f^1$ case\cite{Flouquet2009}.

\begin{figure}[h!]
	\begin{center}
	\includegraphics[width=8cm]{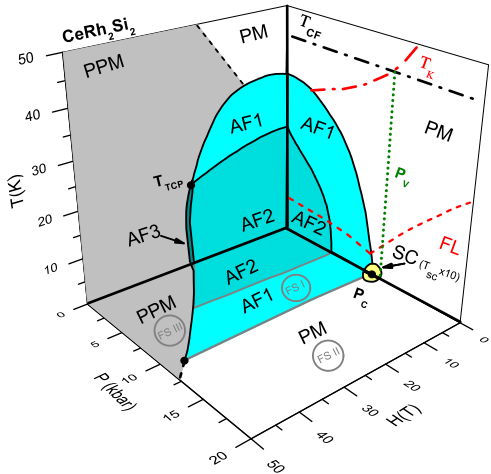}
	\end{center}
	\caption{\label{Fig13} (Color online) (T, H, P) phase diagram of CeRh$_2$Si$_2$ obtained by our TEP measurements and magnetic susceptibility measurements \cite{Knafo2010}: $P_{v}$ is the characteristic valence pressure,  $P_{c}$ is the critical pressure, $T_{CF}$ is the energy of the first cristal field level, $T_{K}$ is the Kondo temperature. Three different Fermi surfaces (labeled FS I, II, III) correspond to the AF, PM, PPM states. Due to the 1$^{st}$ order nature of the quantum phase transition at $P_c$, the FL regime is obeyed bellow the FL temperature shown by the red dashed line. In this regime, resistivity follows AT$^2$ law even at $P_c$.
	}
\end{figure}

\section{Conclusion}
Figure \ref{Fig13} shows the suggested ($T$, $H$, $P$) phase diagram for CeRh$_2$Si$_2$ obtained from TEP measurements. The ($T$, $H$, $P=0$) lines as well as the ($T$, $H=0$, $P$) ones have been determined by TEP measurements. Due to the first order nature of the quantum phase transition at $P_c$, the Fermi liquid (FL) temperature below which $AT^2$ FL law is obeyed is finite at $P_c$ ($T^* \sim$ 2~K). The lines drawn on the  ($T=0$, $H$, $P$) planes (using partly Ref.~\onlinecite{Knafo2010}) are possible guesses with $H_{c}$ terminating at a critical end point with a continuation by a cross over line above $P_c$. Another possibility, as it happens for the system Ce(Rh$_{0.92}$Ru$_{0.08}$ )$_2$Si$_2$ \cite{Machida2013}, will be the collapse of $H_c$ at $P_c$ and thus a decoupling between $H_m$ and $H_c$. In this phase diagram, the superconducting (SC) domain glued on $P_c$ is represented; the SC territory appears quite narrow. Clearly, it is necessary to revisit the SC properties.

Our main message is that AF, PM and PPM phases have three different FS. As with $P$ scan in high magnetic field ($H>H_{c}$ for $P\leqslant P_{c}$ and $H>H_{m}$ for $P > P_{c}$) no phase transition is expected, no $P$ modification of the FS topology may occur in the PPM domain. As discussed for the pseudometamagnetism of CeRu$_2$Si$_2$ \cite{Miyake2006,Daou2006,Suzuki2010}, an itinerant description on CeRh$_2$Si$_2$ close to $P_{c}$ whatever is the strength of the magnetic field seems a sound approach. To confirm the validity of our conclusions, new experiments must be realized. For example at $P=0$, inelastic neutron scattering measurements on single crystal will confirm or not if the intersite magnetic coupling overpasses the Kondo energy and will establish if the crystal field excitations are dispersionless. Further high pressure experiments far above $P_{c}$ will validate or not our proposal of the proximity of $P_{c}$ and $P_{v}$. The loss of the Ising character on crossing $P_{c}$ must be confirmed by  microscopic measurements. By high energy spectroscopy experiments, the pressure evolution of the $4f$ trivalent occupancy number (n$_{f}$) should be determined as well as the crystal structure. A recent advance is a precise determination of the FS at zero pressure in both PM ($T>T_N$) and AF ($T<T_N$) by ARPES \cite{Vyalikh}. Furthermore, dHvA signal change at $H_c$ on entering in the PPM  phase has been recently observed directly \cite{Sheikin2013}.

We thank  H.~Harima, D. Vyalikh, Y. \=Onuki, S. Araki and P. W\"olfle for many useful discussions. This work has been supported by the French ANR (projects PRINCESS), the ERC (starting grant NewHeavyFermion), ICC-IMR, KAKENHI, REIMEI, the EuromagNET II (EU contract no. 228043), LNCMI-CNRS is member of the European Magnetic Field Laboratory (EMFL).

%\bibliographystyle{apsrev4-1}
%\bibliography{biblio}	


\end{document}